\documentstyle[preprint,eqsecnum,aps,psfig]{revtex}

\newcommand{\simgt}{\lower.5ex\hbox{$\; \buildrel > \over \sim \;$}}
\newcommand{\simlt}{\lower.5ex\hbox{$\; \buildrel < \over \sim \;$}}

\begin{document}
\draft
\preprint{HUPD-9909}
\title{Curvature-induced phase transitions in the inflationary universe\\
- Supersymmetric Nambu-Jona-Lasinio Model \\ in de Sitter spacetime -}
\author{J. Hashida, S. Mukaigawa, T. Muta, K. Ohkura and K. Yamamoto}
\address{Department of Physics, Hiroshima University,
         Higashi-Hiroshima, Hiroshima 739-8526}

\date{\today}
\maketitle
\begin{abstract}

The phase structure associated with the chiral symmetry is thoroughly 
investigated in de Sitter spacetime in the supersymmetric 
Nambu-Jona-Lasinio model with supersymmetry breaking terms. 
The argument is given in the three and four space-time dimensions in 
the leading order of the $1/N$ expansion and it is shown that the 
phase characteristics of the chiral symmetry is determined by the 
curvature of de Sitter spacetime. It is found that the symmetry 
breaking takes place as the first order as well as second order 
phase transition depending on the choice of the coupling constant 
and the parameter associated with the supersymmetry breaking term. 
The critical curves expressing the phase boundary are obtained. 
We also discuss the model in the context of the chaotic inflation 
scenario where topological defects (cosmic strings) develop
during the inflation.

\end{abstract}
\vspace{2mm}
\vspace{0.5cm}

\section{Introduction}

In the scenario of the early universe it is understood that
the grand unified theory phase is broken down to the phase of
quantum chromodynamics and electroweak theory through
the Higgs mechanism. While Higgs fields are normally regarded as
elementary fields, it is of interest to consider a possibility that the
Higgs fields may be composed of some fundamental fermions as in the
technicolor model and to see the consequence of this idea in the scenario
of the early universe. On the other hand the supersymmetry is supposed to
be a vital nature possessed by the fundamental unified theory and hence
the incorporation of the supersymmetry in composite Higgs models is
principally important. Under these circumstances it is natural for us
to consider a supersymmetric composite Higgs model in the early stage
of the universe and to see whether any remarkable effects are drawn
during the inflation era.

The Nambu-Jona-Lasinio (NJL) model is a useful prototype model to 
investigate the mechanism of the dynamical symmetry breaking 
\cite{Nam}. In many composite Higgs models the NJL-type Lagrangian 
is employed to realize the dynamical Higgs mechanism. From the 
standpoint of exploring the unified field theory of elementary 
particles it may be of interest to investigate a possibility of a 
supersymmetric version of the NJL model. Unfortunately, however, in 
the supersymmetric version of the NJL model the chiral symmetry is 
strongly protected to keep the boson-fermion symmetry and hence 
the dynamical chiral symmetry breaking does not take place \cite{Lov}. 
If a soft supersymmetry breaking term is added to the 
supersymmetric NJL Lagrangian, the dynamical breakdown of the 
chiral symmetry is brought about for sufficiently large supersymmetry breaking 
parameter $\Delta$ \cite{Ell}.
The reason for this is simple: The large $\Delta$ implies the large 
effective mass of the scalar components of the superfields so that 
quantum effects due to the scalar components get suppressed compared 
with that of the spinor components. Thus the model becomes closer to 
the original NJL model which allows the dynamical fermion mass 
generation. Stating the same substance in a different way we realize 
that the boson by acquiring its mass term forces supersymmetry to 
make balance so that the fermion mass is generated dynamically.

The supersymmetric NJL model with soft supersymmetry breaking is 
useful to study the 
mechanism of the dynamical chiral symmetry breaking within the 
framework of supersymmetry. If we take the model seriously as a 
prototype of the unified field theory, it is natural to extend 
the argument to take into account circumstances of the finite
temperature \cite{HMOA,HMOB} and spacetime-curvature as in the 
early universe \cite{HMOC,IMO,BIO}. 

At the inflation era the quantum effect of the gravitation is of minor
importance while the external gravitational field is non-negligible. Hence we
are naturally led to the supersymmetric NJL model in curved space.
Dealing with the composite Higgs fields is essentially nonperturbative
and does not accept approximate treatments. Accordingly we try to solve the
problem rigorously working in a specific space-time, the de Sitter space,
which possesses a maximal symmetry. The de Sitter space is suitable for
describing the inflationary universe. As a nonperturbative method we rely on
the $1/N$ expansion technique.

The four-fermion interaction model (which is the basis of the NJL model)
in de Sitter space has been discussed by several authors
\cite{Ito,Bu1,IMM,Eli} and is found to reveal the restoration of the
broken chiral symmetry for increasing curvature as a second order phase
transition. The supersymmetric version of the NJL model in curved space
was considered by I. L. Buchbinder, T. Inagaki and S. D. Odintsov \cite{BIO}
in the weak curvature limit. They found that the chiral symmetry is broken
as the curvature increases. Their result is in contrast with the result in
the nonsupersymmetric NJL model. On the other hand the supersymmetric NJL
model in the flat space-time has been investigated by several authors 
\cite{Lov} in the context of dynamical chiral symmetry breaking.

In the present paper we investigate the chiral symmetry breaking phenomena
in the supersymmetric NJL model in de Sitter spacetime induced by the varying
curvature. The situation is considered to be suitable to simulate the phase
transition during the inflationary period.
In the inflationary period the universe expands rapidly with increasing
speed. This phenomenon is often called the de Sitter expansion.
Many investigations on quantum phenomena in de Sitter spacetime 
have been performed motivated by the inflationary paradigm.
The present authors have recently investigated the chiral symmetry 
breaking in the supersymmetric NJL model in de Sitter spacetime in ref. 
\cite{HMOC}. In ref. \cite{HMOC} we have mainly worked in the 
case of three spacetime dimensions. In the present paper we extend the 
previous investigation to the model of four spacetime dimensions, 
and discuss the cosmological consequence of the symmetry breaking phenomenon.

This paper is organized as follows: 
In section 2 we describe our model of the supersymmetric
NJL model in de Sitter spacetime. Here formulae
to obtain effective potentials in de Sitter spacetime in the 1/N
expansion method is described.
In section 3 we consider the case of three spacetime dimensions $D=3$.
The results in section 3 partially overlap the ones in ref.\cite{HMOC}.
We review the phase structure of the chiral symmetry  because the 
case $D=3$ is very instructive for considering the case $D=4$.
The case $D=4$ is investigated in section 4.
In section 5 we consider our model in the context of the chaotic inflation 
and discuss the possible formation of cosmic strings 
during the inflation. Section 6 is devoted to the summary and 
discussions. In this paper we use the units $\hbar=k_B=c=1$, 
and adopt the convention $(-~+~+~+)$ for metric and 
curvature tensors \cite{MTW}.

\section{Supersymmetric Nambu-Jona-Lasinio Model in de Sitter Space}

\subsection{Lagrangian for supersymmetric NJL}

In this section we first summarize the basic ingredients of 
the supersymmetric NJL model in curved spacetime and then we 
evaluate the effective potential in de Sitter spacetime. 
We consider the following Lagrangian for the supersymmetric NJL model
expressed in terms of the component fields of superfields \cite{BIO},
\begin{eqnarray}
  {\cal L} &=& -\nabla^{\mu}\phi^{\dag} \nabla_{\mu}\phi -\rho^2 
  \phi^{\dag}\phi-\nabla^{\mu}\phi^{c \dag} \nabla_{\mu}\phi^{c} 
  -\rho^2 \phi^{c \dag}\phi^{c}
\nonumber
\\
  && -\bar{\psi}(\not{\nabla}+\rho )\psi 
  -\frac{N}{2\lambda}\rho^2, 
\label{LagA}
\end{eqnarray}
where $N$ is the number of components of boson field
$\phi$ and fermion field $\psi$, respectively,
$\lambda$ is the four-fermion coupling constant, and
$\rho$ is the auxiliary field.
Note that in Eq.(\ref{LagA}) only relevant terms in the leading order of 
the $1/N$ expansion are exhibited.

Since the supersymmetry of the model protects the chiral symmetry
from breaking \cite{Lov}, we introduce the following supersymmetry 
breaking terms,
\begin{eqnarray}
  \delta {\cal L}=-\Delta^2\phi^\dagger\phi-\Delta^2\phi^{c\dagger}\phi^c
  -\xi_1 R \phi^\dagger\phi-\xi_2 R \phi^{c\dagger}\phi^c,
\end{eqnarray}
where $R$ is the spacetime curvature, and $\xi_1$, $\xi_2$, and $\Delta$
are coupling parameters, respectively.  
\def\as{{a_{\rm s}}}
\def\bs{{b_{\rm s}}}
\def\cs{{c_{\rm s}}}
\def\afr{{a_{\rm f}}}
\def\bfr{{b_{\rm f}}}
\def\cfr{{c_{\rm f}}}
\def\Tr{{\rm Tr}}
\subsection{Calculation of effective potential}

In this subsection we describe the effective 
potential in curved space. Our strategy is based on the $1/N$ 
expansion and we obtain the effective potential in the leading 
order of the expansion. 
The partition function for the Lagrangian 
$\cal L +\delta \cal L$ in $D$ spacetime dimension is given by
\begin{eqnarray}
  Z&=&{\cal N}\int {\cal D}\rho{\cal D}\phi{\cal D}\phi^{c}
  {\cal D}\psi{\cal D}\bar{\psi} 
  \exp\Bigl[ i\int d^{D}x \sqrt{-g}
\nonumber \\
  &&\times\{\phi^{\dag}(\nabla^{\mu}\nabla_{\mu}-\rho_1^2)\phi
     +\phi^{c \dag} (\nabla^{\mu}\nabla_{\mu} - \rho_2^2 )\phi^{c} 
\nonumber \\
  &&\hspace{1cm}-\bar{\psi}(\not{\nabla}+\rho )\psi
    -\frac{N}{2\lambda}\rho^2  \} \Bigr],
\end{eqnarray}
up to the normalization constant, where $\rho_1$ and $\rho_2$
are defined by
\begin{equation}
  \rho_j^2=\rho^2+\Delta^2+\xi_j R,
\label{defrhoi}
\end{equation}
with $j=1,2$. Integrating over 
$\phi$,~$\phi^c$,~$\psi$, and $\bar\psi$, we find
\begin{eqnarray}
  Z&=&\int {\cal D}\rho 
  [ det(\nabla^{\mu}\nabla_{\mu} - \rho_1^2 )]^{-N}
  [ det(\nabla^{\mu}\nabla_{\mu} - \rho_2^2 )]^{-N}
\nonumber 
\\
 && \times [det(\not{\nabla} + \rho )]^{N}\exp\Bigl[i\int d^{D} x 
  \sqrt{-g}(-\frac{N}{2\lambda}\rho^2 ) \Bigr] 
\nonumber 
\\
 &=&\int {\cal D}\rho 
  \exp\Bigl[i\int d^{D}x \sqrt{-g} \Bigl(-\frac{N}{2\lambda}\rho^2\Bigr) 
\nonumber 
\\
  && -N 
  \ln det(\nabla^{\mu}\nabla_{\mu} - \rho^2-\Delta^2-\xi_1 R ) 
\nonumber 
\\
  && -N 
  \ln det(\nabla^{\mu}\nabla_{\mu} - \rho^2-\Delta^2-\xi_2 R ) 
\nonumber 
\\
  && +N \ln det(\not{\nabla} + \rho) \Bigr]. 
\end{eqnarray}
The effective action for Large $N$ is written as
\begin{eqnarray}
\Gamma[\rho] 
&=& 
S_{eff}[\rho] + O(\frac{1}{N}) ,
\end{eqnarray}
and the effective potential $V(\rho)$ in the leading order of the
$1/N$ expansion can be explicitly calculated such that
\begin{eqnarray}
  V(\rho)&=& 
  -\frac{S_{eff}[\rho = \mbox{const}]}{\Omega} \nonumber \\
 &=&
  - \frac{1}{\Omega} \Bigl[
  \int d^{D}x \sqrt{-g} (-\frac{\rho^2}{2\lambda})
\nonumber
\\
    &&   -i\ln \frac{det(\not{\nabla} + \rho )}{det(\not{\nabla})}
\nonumber
\\
    &&   +i\ln \frac{det(\nabla^{\mu}\nabla_{\mu}-\rho^2-\Delta^2-\xi_1R )}
                     {det(\nabla^{\mu}\nabla_{\mu}-\Delta^2-\xi_1R)}
\nonumber
\\
    &&   +i\ln \frac{det(\nabla^{\mu}\nabla_{\mu}-\rho^2-\Delta^2-\xi_2R )}
                     {det(\nabla^{\mu}\nabla_{\mu}-\Delta^2-\xi_2R)}
          \Bigr]
\nonumber
\\
&=&
  -\frac{1}{\Omega} \Bigl[
  \int d^{D}x \sqrt{-g} (-\frac{\rho^2}{2\lambda})
\nonumber
\\
    &&    -i \{ Tr\ln(\not{\nabla}+\rho) - Tr\ln(\not{\nabla}) \}
\nonumber
\\
    &&    +i \{ Tr\ln(\Box -\rho^2-\Delta^2-\xi_1R ) 
\nonumber
\\
    &&\hspace{1cm}         - Tr\ln(\Box-\Delta^2-\xi_1R)\} 
\nonumber
\\
    &&    +i \{ Tr\ln(\Box -\rho^2-\Delta^2-\xi_2R )
 \nonumber
\\
    &&\hspace{1cm}        - Tr\ln(\Box-\Delta^2-\xi_2R)\}\Bigr] ,
\label{Vrho}
\end{eqnarray}
where  $\Omega$ is the spacetime volume and  
$\Box$ denotes $\nabla^{\mu}\nabla_{\mu}$. 
The effective potential (\ref{Vrho}) is normalized so that $V(0)=0$.
Eq.(\ref{Vrho}) may be rewritten in the form,
\begin{eqnarray}
V(\rho)&=& \frac{\rho^2}{2\lambda} 
     +i \int^{\rho}_{0}   ds ~tr \big<x|(\not{\nabla} + s)^{-1}|x\big>
\nonumber
\\
    &&+i \int^{\rho^2}_{0} dt~\big<x|(\Box -t-\Delta^2-\xi_1R)^{-1}|x\big>
\nonumber
\\
    &&+i \int^{\rho^2}_{0} dt~\big<x|(\Box -t-\Delta^2-\xi_2R)^{-1}|x\big>.
\label{EP}
\end{eqnarray}
It is important to note that the last three terms on the right-hand side 
of Eq.(\ref{Vrho}) are related to the massive boson and fermion propagators
$S(x,y;s)$ and $G(x,y;t)$. 
If we define the functions, $S(x,y;s)$ and $G(x,y;t)$, which satisfy
\begin{eqnarray}
  ({\ooalign{\hfil/\hfil\crcr$\nabla$}}+s)S(x,y ;s)
  &=& 
  \frac{\delta^{(D)}(x,y)}{\sqrt{-g(x)}}, 
\\
  (\Box -t )G(x,y;t)
  &=&
  \frac{\delta^{(D)}(x,y)}{\sqrt{-g(x)}},
\end{eqnarray}
respectively, it is easy to find the relationship :
\begin{eqnarray}
  S(x,y ;s) &=& \big<x|(\not{\nabla} + s)^{-1}|y\big>,
\\
  G(x,y ;t) &=& \big<x|(\Box -t)^{-1}|y\big>,
\end{eqnarray}
by making use of the following equality
\begin{eqnarray}
&&
  \int d^{D}x \sqrt{-g(x)}\; |x\big>\big<x| = \hat{\mbox{\boldmath{1}}},
\\&&
  \big<x|y\big> = \frac{\delta^{(D)}(x,y)}{\sqrt{-g(x)}}.
\end{eqnarray}
Thus the calculation of the effective potential 
in the leading order of the $1/N$ expansion reduces to the 
evaluation of the propagators in the de Sitter spacetime,
\begin{eqnarray}
  V(\rho)&=&{\rho^2\over 2\lambda}+i\int_0^\rho ds\Tr S(x,x;s)
\nonumber
\\
  &&+i\int_{0}^{\rho^2} dt G(x,x;t+\Delta^2+\xi_1 R)~
\nonumber
\\
  &&+i\int_{0}^{\rho^2} dt G(x,x;t+\Delta^2+\xi_2 R)~.
\label{VgenA}
\end{eqnarray}
The next task is to write down the fermion and the boson 
propagators in de Sitter spacetime. 
\subsection{Boson and fermion propagators in de Sitter spacetime}
In this subsection we briefly review the boson and fermion 
propagators in de Sitter spacetime.
The de Sitter spacetime is defined as the maximally symmetric
curved spacetime. The quantum field theory in de Sitter 
spacetime has been studied extensively and the propagator for scalar 
fields is well-known \cite{BirrellDavies,BunchDavies,Allen,AllenJ,Tagirov}.
The scalar propagator with mass squared $t$ in the D dimensional 
de Sitter spacetime is explicitly written as 
\begin{equation}
  G(x,y;t)=-i{r^{2-D} \over (4\pi)^{D/2}}
  {\Gamma(\as)\Gamma(\bs)\over\Gamma(\cs)}
  {}_2F_1(\as,\bs,\cs;1-z),
\label{BGF}
\end{equation}
where $\Gamma(a)$ is the gamma faction, ${}_2F_1(a,b,c;1-z)$ is the 
hypergeometric function, and
\begin{eqnarray}
  &&a_s={1\over2}\biggl(D-1+\sqrt{(D-1)^2-4tr^2}\biggr),
\\
  &&b_s={1\over2}\biggl(D-1-\sqrt{(D-1)^2-4tr^2}\biggr),
\\
  &&c_s={D\over2},
\end{eqnarray}
with
\begin{eqnarray}
  &&z={\sigma^2\over 2r^2}, \hspace{1cm}
  \sigma^2={1\over2}(\vec x-\vec y)^2.
\label{geodesicd}
\end{eqnarray}
Note that $\sigma$ is in proportion to the geodesic distance
between the two points $x$ and $y$, and $r$ is the radius of de Sitter space, 
which is related to the Hubble parameter by $H=1/r$. 
It should be emphasized that the non-minimal coupling term gives
effective mass $ D(D-1) \xi/r^2$, where $R=D(D-1)/r^2$.

The fermion propagator is written as \cite{Mukaigawa,Camporesi}
\begin{equation}
  S(x,y;s)=\bigl(A(x,y;s)+B(x,y;s)\sigma_{;\mu}\gamma^\mu\bigr)U,
\end{equation}
where U is the matrices composed of the Dirac matrices and 
\begin{eqnarray}
 A(x,y;s)&=&i{sr^{2-D} \over (4\pi)^{D/2}}
  {\Gamma(a_f)\Gamma(a_f^*)\over\Gamma(c_f)}
\nonumber
\\ 
  &&\times\sqrt{1-z}{}_2F_1(a_f,a_f^*,c_f;1-z),
\label{Axy}
\end{eqnarray}
where
\begin{eqnarray}
  &&a_f={D\over2}+ isr, \hspace{0.5cm}c_f={D\over2}+1.
\end{eqnarray}
We do not present the explicit expression of the invariant
function $B(x,y;s)$ because it is irrelevant to our purpose
of calculating the effective potential. In fact we find
\begin{eqnarray}
  \Tr S(x,x;s)&=&\lim_{z\rightarrow0}A(x,y;s)\Tr U
\nonumber
\\  &=&\Tr[1]\lim_{z\rightarrow0}A(x,y;s),
\end{eqnarray}
with the normalization $\lim_{z\rightarrow0}\Tr U=\Tr[1]$.

Now we are ready to calculate the effective 
potential (\ref{VgenA}). It is instructive to
investigate the case $D=3$ \cite{HMOC},
because the effective potential can be written 
in terms of elementary functions.
We first review the phase structure of the case $D=3$
in the next section, and then the case $D=4$
is studied in section 4.
\section{Analysis in $3$ Dimensions}
\subsection{Gap Equation}

Following the previous argument, the expression for 
the effective potential is given by 
\begin{eqnarray}
  V(\rho)&=&{\rho^2\over 2\lambda}+i\int_0^\rho ds\Tr S(x,x;s)
\nonumber
\\
  &&+2i\int_{\Delta^2+\xi R}^{\Delta^2+\xi R+\rho^2} dt 
  G(x,x;t)~,
\label{Vgen}
\end{eqnarray}
where we set $\xi_1=\xi_2=\xi$ for simplicity.

In order to write down the effective potential we consider
the coincidence limit of the propagator, $lim_{y\rightarrow x} G(x,y;t)$. 
In the coincidence limit, the propagator diverges in general.
Then we adopt the point splitting method and consider
$G(x,y;t)$ around $z=0$. For that purpose we use the mathematical 
formula (e.g., \cite{Magnus}):
\begin{eqnarray}
  &&{\Gamma(a)\Gamma(b)\over\Gamma(c)}
  {}_2F_1(a,b,c;1-z)=
\nonumber
\\
  &&\hspace{0.cm}\Gamma(a+b-c)z^{c-a-b}{}_2F_1(c-a,c-b,c-a-b+1;z)
\nonumber
\\
  &&\hspace{0.cm}+{\Gamma(a)\Gamma(b)\Gamma(c-a-b)\over 
  \Gamma(c-a)\Gamma(c-b)}{}_2F_1(a,b,a+b-c+1;z)~.
\end{eqnarray}
With the use of this formula the boson propagator is written as
\begin{equation}
  G(x,y;t)={-i\over 8\pi rz^{1/2}}+{i\over 4\pi r} \nu {\rm coth} \pi\nu 
  +{\cal O}(z^{1/2})~,
\label{Gthree}
\end{equation}
where we defined $\nu=\sqrt{tr^2-1}$. Analytic continuation is
needed for $tr^2<1$.
On the other hand for $z\rightarrow0$ Eq.(\ref{Axy}) reduces to
\begin{eqnarray}
  A(x,y;s)&=&{is\over 8\pi rz^{1/2}}
  -{is\over 4\pi r}\biggl({1\over4}+s^2 r^2\biggr) 
  {{\rm tanh} \pi s r\over s r}
\nonumber
\\  &&+{\cal O}(z^{1/2})~.
\label{Athree}
\end{eqnarray}

Inserting (\ref{Gthree}) and (\ref{Athree}) into (\ref{Vgen}),
then the effective potential turns out to be,
\begin{eqnarray}
  V(\rho)&=&{\rho^2\over 2\lambda}
  +{1\over \pi r}\int_{0}^{\rho} 
  ds s  \biggl({1\over 4}+s^2 r^2\biggr)
  {\tanh \pi s r\over sr}
\nonumber
\\
  &&\hspace{0.0cm}
  -{1\over 2\pi r}\int_{\Delta^2+\xi R}^{\Delta^2+\xi R+\rho^2} dt \nu
  {\rm coth}\pi\nu,
\end{eqnarray}
where we have set $\Tr[1]=4$.\footnote{
For $D=3$ we adopt the reducible representation of the Clifford
algebra of Dirac matrices in order to guarantee the existence of 
$\gamma_5$.}
Note that the divergent terms cancel each other, which would 
originate from the supersymmetry.

The gap equation, ${\partial V/\partial (\rho^2)}=0$, reads
\begin{eqnarray}
  &&{\pi r\over \lambda}
  +\biggl({1\over4}+\rho^2r^2\biggr){\tanh\pi \rho r \over \rho r}
\nonumber
\\  &&-\sqrt{\rho^2r^2+6\zeta-1}
  {\rm coth\pi\sqrt{\rho^2r^2+6\zeta-1}}=0~,
\end{eqnarray}
where we defined 
\begin{equation}
  \zeta=\xi+{1\over 6}\Delta^2 r^2.
\end{equation}
Note that the solution of this gap equation is completely specified
by two parameters $\zeta$ and $\pi r/\lambda$. This means that
the phase structure derived from the effective potential is completely
specified in the plane of these two parameters. 

\subsection{Phase Structure}
We present the typical shape of our effective potential 
in the case $D=3$ in Figure 1 (see also \cite{HMOC}). 
In the region labeled by 'S', the effective potential behaves so that
the symmetry is unbroken. In the region labeled by 'B', the symmetry 
is broken through the second order phase transtion.
Finally in the region labeled by 'F', the symmetry 
is broken through the first order phase transtion.
The boundaries in Fig.~1, which separate the phases, 
are obtained by direct observation of numetical analysis
of the effective potential. Analytically the condition
is described as follows. The solid line in Fig.~1 is found by solving
\begin{equation}
  {\partial V \over \partial (\rho^2)}\Bigg|_{\rho=0}=0,
\end{equation}
which is explicitely written as
\begin{equation}
  {\pi r\over \lambda}+{\pi\over 4}-\sqrt{6\zeta-1}\coth\pi
\sqrt{6\zeta-1}=0.
\label{d1V3}
\end{equation}
On the other hand, the condition for the dashed line 
is rather complicated. The condition is $V(\rho_*)=0$,
where $\rho_*={\rm max}\{\rho_1,\rho_2\}$ when the equation
${\partial V / \partial (\rho^2)}=0$ has two different solutions
$\rho_1$ and $\rho_2$.

The branching point in Fig.~1 is of special interest. It is a critical
point which divides the broken phase into type 'F' and type 'B'. 
At the branching point C the following conditions are found to be 
satisfied simultaneously:
\begin{equation}
{\partial V \over \partial (\rho^2)}\Bigg|_{\rho=0}=0, \ \ \ 
{\partial^2 V \over \partial (\rho^2)^2}\Bigg|_{\rho=0}=0. \label{d2V3}
\end{equation}
The conditions are explicitly given respectively by Eq.~(\ref{d1V3}) and
\begin{equation}
 1-\frac{\pi^2}{12}+\frac{1}{2\sinh^2\pi\sqrt{6\zeta-1}}
 -\frac{\coth\pi\sqrt{6\zeta-1}}{2\pi\sqrt{6\zeta-1}}=0. 
 \label{Crit3}
\end{equation}
{}From Eq.~(\ref{Crit3}) we find that $\zeta=0.290138(=:\zeta_*)$ 
at the branching point. By substituting this value for $\zeta$ 
in Eq.~(\ref{d1V3}) we obtain $\pi r/\lambda=0.08306$ at the 
branching point.

Since we have seen the phase structure of our effective potential,
we now discuss the time-evolution of the chiral symmetry 
assuming an inflation background. This consideration
will make sense if the Hubble parameter (radius of de Sitter 
spacetime) changes slowly and 
the effective potential in the inflation phase is well described 
by that in de Sitter spacetime found above. 
As we will see in section 5, the Hubble parameter slowly decreases 
as the universe evolves during the chaotic inflation in actual.
And the curvature of de Sitter spacetime slowly decreases
(radius $r$ increases). 

First let us consider the case $\Delta=0$.
In this case $\zeta$ is equal to $\xi$. If the coupling constant
$\lambda$ and $\xi$ are fixed, we move from left to right
by increasing $r$ in Fig.~1. This indicates that the chiral symmetry will 
be ultimately restored for sufficiently large $r$.
The initial situation depends on the value of $\xi$.
By direct observation of the gap equation one can easily
show that if the parameter $\xi$ is kept below $1/4$ 
the effective potential stays in the region 'S' in the limit
of $\pi r/\lambda=0$. Thus the phase transition does not occur
for $\xi<1/4$. On the other hand the effective potential is in 
a broken phase of type 'F' or 'B' if we take the value of $\xi$ 
above $1/4$. Then the phase restoration occurs as the curvature 
decreases.

Second we consider the case $\xi=0$ and $\Delta\neq0$.
In this case $\zeta=\Delta^2r^2/6$. 
If the coupling constant $\lambda$ and $\xi$ are fixed,
then the trajectory for varying $r$ on the $(\zeta,\pi r/\lambda)$ 
plane will be the one as shown in Fig.2.
By increasing the radius $r$, we move 
from the bottom to the top along the straight line. For each line,
the value of $\Delta \lambda/\pi$ is fixed (a) $\Delta \lambda/\pi=100$,
(b) $\Delta \lambda/\pi=15.9$, and (c) $\Delta \lambda/\pi=2$.
Along the curve for $\Delta \lambda/\pi=100$
a first order phase transition occurs, however, 
a second order transition occurs for $\Delta \lambda/\pi=2$. 
The critical case is $\Delta \lambda/\pi=15.9$. 
For $\Delta \lambda/\pi< 1$ the phase transition does not occur.

Thus in the model $\Delta\neq0$ and $\xi=0$, 
the phase of the symmetry is in the symmetric phase first, and
it is broken as the curvature of the universe decreases as long as
$\Delta \lambda/\pi\geq 1$.
On the other hand the symmetry is ultimately unbroken
in the case $\Delta=0$.
This case is not relevant to our conventional idea of 
symmetry breaking.
Thus we focus our attention to the case $\Delta\neq0$ and $\xi=0$ in the 
following arguments.

\section{Analysis in 4 Dimensions}
\subsection{Gap Equation}
In this section we consider the case $D=4$. 
From Eq.(\ref{VgenA}) or (\ref{Vgen}) the gap equation, 
$\partial V/\partial (\rho^2)=0$, reads, 
\begin{eqnarray}
  {1\over 2\lambda}+{i\over 2\rho}{\rm Tr} S(x,x;\rho)
  +2i G(x,x;\rho^2+\Delta^2)=0.
\end{eqnarray}
Here we consider the case $\Delta\neq0$ and 
$\xi_1=\xi_2=0$, because this choice of parameter 
is relevant to the conventional idea of symmetry breaking phenomena
as described in section 3.

The boson propagators (\ref{BGF}) in the case $D=4$ reduces to
\footnote{In this section we use the Hubble parameter $H(=1/r)$ instead of
the radius $r$ to describe the curvature of de Sitter spacetime.}:
\begin{eqnarray}
  &&G(x,y;t)= {-i H^2 \over (4\pi )^{2}}
  {\Gamma({3/2}+\nu)\Gamma({3/2}-\nu)\over\Gamma(2)}
\nonumber
\\
  &&\hspace{1cm}\times  {}_2F_1(3/2+\nu,3/2-\nu,2;1-z),
\end{eqnarray}
where we defined
\begin{eqnarray}
  &&\nu=\sqrt{{9\over4}-{t\over H^2}-{\Delta^2\over H^2}}.
\end{eqnarray}
While the function $A$ of Eq.(\ref{Axy}) is
\begin{eqnarray}
  &&A(x,y;s) = {is H^2 \over (4\pi)^{2}}
  {\Gamma(2+is/H)\Gamma(2-is/H)\over\Gamma(3)}
\nonumber
\\
  &&\hspace{1cm}\times\sqrt{1-z}{}_2F_1(2+is/H,2-is/H,3;1-z).
\end{eqnarray}

Since the propagators diverges in the coincidence limit, 
$z\rightarrow0$, we adopt the point splitting regularization
by keeping $z$ small but finite.
In this case it is useful to use the mathematical formula 
(e.g., \cite{Magnus})
\begin{eqnarray}
&&{\Gamma(a)\Gamma(b)\over\Gamma(a+b-1)}
   {}_2F_1(a,b,a+b-1;1-z)=
\nonumber
\\
&&{1\over z}+(a-1)(b-1)\sum_{n=0}^\infty
  {(a)_n(b)_n\over n!(n+1)!} \Bigl\{\ln(z)+\psi(a+n)
\nonumber
\\
  &&\hspace{1cm}
  +\psi(b+n)-\psi(n+1)-\psi(n+2)\Bigr\}z^n,
\end{eqnarray}
where $(a)_n=\Gamma(a+n)/\Gamma(a)$ and $\psi(x)$ is the
polygamma function. With the use of this formula the gap 
equation reduces to
\begin{eqnarray}
  &&{(2\pi /H)^2\over \lambda}+{1-\sqrt{1-z}\over z}
\nonumber
\\  
  &&-\sqrt{1-z}(1+\rho^2/H^2)\sum_{n=0}^\infty
   {(2+i\rho /H)_n(2-i\rho /H)_n\over n!(n+1)!} 
\nonumber
\\
  &&\hspace{0.cm} 
    \times\Bigl\{\ln(z)+\psi(2+i\rho /H+n)
  +\psi(2-i\rho /H+n)
\nonumber
\\
  &&\hspace{1.cm} -\psi(n+1)-\psi(n+2)\Bigr\}z^n
\nonumber
\\
  &&\hspace{0.cm} 
  +(-2+\rho^2 /H^2+\Delta^2/H^2)\sum_{n=0}^\infty
   {(3/2+\nu)_n(3/2-\nu)_n\over n!(n+1)!} 
\nonumber
\\
  &&\hspace{0.cm} 
    \times\Bigl\{\ln(z)+\psi(3/2+\nu+n)+\psi(3/2-\nu+n)
\nonumber
\\
  &&\hspace{1.cm} 
  -\psi(n+1)-\psi(n+2)\Bigr\}z^n=0.
\label{gapfour}
\end{eqnarray}
The divergent terms of order ${\cal O}(z^{-1})$, which appears
in the boson and fermion propagators, cancel each other
in the effective potential.
This would be traced back to the supersymmetry of the theory.
However the logalithmic divergent terms ${\cal O}(\ln z)$
arise in the case $D=4$. Thus our theory depends on the
cutoff parameter. Instead of $z$ we introduce the
momentum cutoff parameter $\Lambda$. We can show that the geodesic 
distance $\sigma=(\vec x-\vec y)^2/2$ is related to the
momentum cutoff parameter $\Lambda$ such that $\sigma=\Lambda^{-2}$.
Then we have the relation
\begin{equation}
z={H^2\over 2 \Lambda^2},
\end{equation}
where we used Eq.(\ref{geodesicd}).

\subsection {Phase structure}

Next we consider the phase structure of the model by solving the
gap equation (\ref{gapfour}). Because of the complexity of 
equation, an analytic approach would not be useful in the case $D=4$.
Then we adopt numerical method by calculating equation (\ref{gapfour}).
As the gap equation depends on the cutoff parameter, we do not find 
a simple scaling relation in gap equation (\ref{gapfour}), 
on the contrast to the case $D=3$. Therefore we fix the cutoff 
parameter $\Lambda=10^{15}{\rm GeV}$ in the following arguments, 
for simplicity. 

Fig.~3 shows the phase structure of the effective potential 
on the $(H/\Lambda,1/\lambda\Lambda^2)$-plane. 
The curves in the figure show the phase boundary between a 
symmetric phase and a broken phase for $\Delta=$
$10^{15}{\rm GeV},~3\times10^{14}{\rm GeV},~10^{14}{\rm GeV}$, 
respectively. In the right and upper region of each curve
in Fig.~3, the effective potential behaves so that 
the symmetry is unbroken.
On the other hand in the left and lower region of
each curve the symmetry is broken.
We discuss the type of the broken phase, i.e., first order or
second order, in the later.

By decreasing the curvature (decreasing $H$) 
with the coupling constants, $\Lambda$, $\Delta$, and $\lambda$ fixed,
we move from right to left in Fig.~3. This indicates that
the effective potential is in the symmetric phase initially, 
however, the symmetry is broken as $H$ decreases 
if the coupling constant $\lambda$ is larger then a critical value.
Let us find the critical value of the coupling constant, which 
depends on the supersymmetry breaking mass 
$\Delta$ and the cutoff parameter $\Lambda$.
Taking the limit $H\rightarrow 0$ of Eq.(\ref{gapfour}), we find the 
following critical coupling constant $\lambda_{\rm cr}$,
\begin{eqnarray}
  &&{1\over \lambda_{\rm cr}}=-{\Delta^2\over(2\pi)^2}\sum_{n=0}^\infty
  {1\over n!(n+1)!}
\nonumber
\\  
  &&\times \biggl[\ln {\Delta^2\over2\Lambda^2}--\psi(n+1)-\psi(n+2)\biggr]
  \bigg({\Delta^2\over2\Lambda^2}\biggr)^n.
\label{criticall}
\end{eqnarray}
We show the critical value $\lambda_{\rm cr}$ 
as a function of $\Delta/\Lambda$ in Fig.~4. 
We see the expected result in the figure that the critical 
coupling constant $\lambda_{\rm cr}$ must becomes larger as the 
the supersymmetry breaking mass $\Delta$ becomes smaller.

Finally in this section we discuss the effective potential
of type of the first order phase transition. 
Similar to the case $D=3$ we find that the effective potential 
behaves so that the phase transtion occurs through first order 
transition for a specific range of coupling constants.
The region is small and we can not recognize it in Fig.~3.
Fig.~5 shows the region in Fig.3 under magnification for demonstraition,
where we adopted the case $\Delta=10^{15}{\rm GeV}$.
In the narrow region labeled by 'F' the effective potential
has the shape of the first order phase transition. 
For other cases, $\Delta=3\times 10^{14}{\rm GeV}$ and 
$\Delta=10^{14}{\rm GeV}$ in Fig.~3,
we find narrow regions of 'F' in a similar way.
Thus the model for $D=4$ has the similar phase structure 
as the case $D=3$.

Because there appears many parameters in the case $D=4$ and we can not 
find a good  scaling relation on the contrast to the case $D=3$. 
Due to this fact it is difficult to show the phase structure in 
case $D=4$ clearly. As will be discussed in the next section 
the change of the Hubble parameter is rather fast in the
chaotic inflation model. Then the period that the
the effective potential has the shape of the first order 
phase transition is very short compared with the Hubble 
expansion time. Then the cosmological importance of the
first order transition is unclear at present. Therefore we only
focus on the epoch of the (second or first) phase transtion
in the inflationary universe and discuss other cosmological 
consequences because phase transitions
in the universe predicts topological defects in general.

\section{Epoch of phase transtion in the chaotic inflationary model}
\def\as{{a_{\rm s}}}
\def\bs{{b_{\rm s}}}
\def\cs{{c_{\rm s}}}
\def\afr{{a_{\rm f}}}
\def\bfr{{b_{\rm f}}}
\def\cfr{{c_{\rm f}}}
\def\Tr{{\rm Tr}}
\def\phis{{\rho}}
\def\hphis{{\hat\rho}}
\def\hDelta{{\hat\Delta}}
\def\as{{a_{\rm s}}}
\def\bs{{b_{\rm s}}}
\def\cs{{c_{\rm s}}}
\def\afr{{a_{\rm f}}}
\def\bfr{{b_{\rm f}}}
\def\cfr{{c_{\rm f}}}
\def\Tr{{\rm Tr}}
\def\Mpl{M_{\rm PL}}
In this section we consider a cosmological application
of our model in the chaotic inflation universe \cite{Linde}.
Cosmological phase transition predicts
the formation of topological defects in general. 
In our model $U(1)$ chiral symmetry is broken down and 
the formation of cosmic strings would be predicted by the phase 
transtion. In general topological defects are harmful, 
as examplified by the monopole problem and domain wall problem.
However, cosmic strings have been studied with the motivation
of the cosmic structure formation. 
The observational confrontation of the cosmic 
string scenario is still at issue\cite{HMa}.

The investigations in the present paper show that the symmetry 
breaking may occur during inflation era, which is
triggered by the change of the curvature of spacetime. 
As an application we consider the phase transition in the context
of the chaotic inflation model and the formation
of the cosmic strings during the inflation.
If the inflation lasts sufficiently long after the 
phase transition, the defects will be diluted.
However, if the phase transition occurs at a
suitable epoch of inflation, the dilution of the strings
is not completed. Similar investigations were given on the 
basis of a simple model of curvature induced phase transition
\cite{Yokoyama,Lyth}. 
Following their consideration 
we here discuss the epoch of the formation of the cosmic strings 
in our model during the chaotic inflation era.

We assume the chaotic inflation background.
To be specific we assume that the inflation 
is derived by the field $\phi$ with the 
Lagrangian
\begin{equation}
  {\cal L}_\phi=-\nabla^\mu \phi \nabla_\mu \phi
  -{1\over 2} m^2\phi^2.
\end{equation}
Here we set  $m= 10^{13} {\rm GeV}$ \cite{Lyth}, which 
is the constraint from the amplitude of density 
perturbations for a successful inflation scenario. 
With the background metric, 
\begin{equation}
  ds^2=-dt^2+a(t)^2 d{\bf x}^2,
\end{equation}
the equations of motion under the 
slow roll approximation take the form,
\begin{eqnarray}
  &&3H\dot\phi+m^2\phi=0,
\\
  &&H^2={4\pi \over 3 \Mpl^2}{m^2\phi^2},
\label{defH}
\end{eqnarray}
where $H=\dot a/a$ is the hubble parameter, 
and the dot denotes (cosmic time) $t$ differentiation. 
The solution of inflation is well known
\begin{eqnarray}
  &&\phi(t)=\phi_i-{m\Mpl\over 2\sqrt{3\pi}}(t-t_i),
\\
  &&a(t)=a_i \exp \Biggl[{2\pi\over \Mpl^2} (\phi_i^2-\phi(t)^2)\Biggr],
\\
  &&H(t)=\sqrt{4\pi m^2\over 3\Mpl^2}\phi(t),
\\
  &&R=12H^2+6 \dot H={16\pi \over\Mpl^2} m^2\phi^2(t)-2m^2
\nonumber
\\ &&\hspace{0.4cm}\simeq{16\pi \over\Mpl^2} m^2\phi^2(t),
\end{eqnarray}
where $\Mpl$ is the Plank mass, and the subscript `i' denotes 
the value at the initial time when the inflation started.

In this chaotic inflation model the horizon crossing 
(during the inflation) of the fluctuation of
the wave length corresponding to the present horizon size 
occurs at $\phi(t)\sim 3\Mpl$. 
The epoch of forming cosmic string are estimated as follows.
The phase of $\rho$  does not fix soon after the phase 
transition because of the quantum fluctuations of the $\rho$ 
field. Following the previous investigations on curvature 
induced phase transition \cite{Yokoyama,Lyth,Linde}, 
the phase of $\rho$ fixes at the time, when the condition
is satisfied
\begin{equation}
  H^2=-\lim_{\rho\rightarrow 0}V(\rho)''~,
\label{stringform}
\end{equation}
which we may regard as the time of the string formation 
during the inflation.
We denote the value of $\phi(t)$ at this time as $\phi_2$. 
With the use of equation (\ref{defH}), $\phi_2$ is written as
\begin{equation}
  \phi_2=\sqrt{-{3\Mpl^2\over4\pi m^2}
  \lim_{\rho\rightarrow 0}V(\rho)''}.
\label{phini}
\end{equation}

In the previous work \cite{Yokoyama,Lyth} 
the formation of the cosmic strings during inflation was discussed 
to avoid observational difficulties of the cosmic string scenario.
They argued the range of the model parameter (coupling constant) 
to realize the suitable formation of the cosmic strings.
In this paper we consider the observational possibility
and find the condition that the cosmic strings are formed 
during the inflation without complete dilution of the cosmic strings 
so that the cosmic strings may exist in the universe at present time.
The condition is simply given by 
\begin{equation}
  {\Mpl\over\sqrt{4\pi}}\simlt\phi_2\simlt{ 3\Mpl},
\label{condition}
\end{equation}
where we have assumed that the strings are formed at the rate 
of one string per horizon volume at the formation epoch.

This condition constrains the coupling constant 
$\lambda$ in our model. From Eq.(\ref{stringform}), 
we have 
\begin{eqnarray}
  &&{1\over \lambda}+H^2-{H^2\over 
  (2\pi)^2}\sqrt{1-z}{}_2F_1(2,2,3;1-z)
\nonumber
\\  
  &&+{H^2\over (2\pi)^2}{\pi (1/4-\mu^2)\over \cos\pi\mu}
  {}_2F_1(3/2+\mu,3/2-\mu,2;1-z)=0,
\nonumber
\\ 
\end{eqnarray}
where $\mu=\sqrt{9/4-\Delta^2/H^2}$, 
$z=H^2/2\Lambda^2$, and $H$ is specified as
\begin{equation}
  H=\sqrt{4\pi m^2\over3\Mpl^2}\phi_2.
\end{equation}

Fig.6 shows the formation time of the cosmic strings on the parameter 
space, $\lambda$ and $\Delta$. Here we set $\Lambda=10^{15}{\rm GeV}$.
The result depends on the choice of cutoff parameter $\Lambda$. However
the qualitative feature does not depend on the choice of $\Lambda$.
The region between the two solid curves in 
Fig.6 satisfies the condition (\ref{condition}). The lower solid curve
is from  the condition that the formation of the strings occurs at 
$\phi_2=3\Mpl$. While the upper solid curve comes from 
$\phi_2=\Mpl/\sqrt{4\pi}$.
The larger coupling constant $\lambda$ leads to the earlier formation 
of the strings during the inflation. And the larger breaking parameter 
$\Delta$ also leads to the earlier formation of the strings. Fig.6 shows this
expected result. Thus if we make a choice of parameters in the
lower region than the lower solid curve, the strings
are diluted out by the inflation and they would not be observable.
\footnote{If the reheating temperature was so high that 
the symmetry restored by the finite temperature effect, 
the cosmic string would be 
produced again.} The dashed line, which is almost overlapped with the
upper solid line, shows the critical coupling constant 
for the phase transitions (see also Fig.~4).
Therefore the strings do not form when we make a choice of 
parameters in the region above the dashed line. 

\section{Conclusions}
In summary we have shown that the curvature of spacetime might 
be a trigger of phase transition during the inflation in a class 
of model of the dynamical symmetry breaking. 
To be specific we have investigated the phase structure 
of the supersymmetric NJL-model in de Sitter spacetime.
By evaluating the effective potential in the leading
order of the $1/N$ expansion, we have examined the 
phase structure of the chiral symmetry in three- and 
four-spacetime dimensions.
We have investigated how the curvature of de Sitter spacetime
changes the phase structure, and we have found that the 
symmetry breaking takes places as the first order as well 
as second order phase transition depending on the coupling
constant and the parameter of the supersymmetry breaking in both
cases of three- and four-spacetime dimensions.
\footnote{In the case of the potential barrier is 
small enough, the conventional picture of
first order phase transition would be broken down 
because of the quantum fluctuation of the field.}

In the case of three-spacetime dimensions, the 
gap equation reduces to a simple equation written
by elementary functions. Divergent terms cancel each other
and we do not need introduce a cutoff parameter.
As shown in section 3, the shape of the effective potential can 
be specified by the two parameters, $r/\lambda$ and $\zeta$, 
which characterizes the phase structure.
This model might not have physical meaning, however, it is very 
instructive to consider the case of four-spacetime dimensions.

In the case of four-spacetime dimensions,
the gap equation does not reduce to a simple form,
which is written by the hypergeometric functions.
We have to introduce a (momentum) cutoff parameter to regularize 
the theory. The analysis is tedious, however, we have found that
the phase structure is very similar to the case of
three-spacetime dimensions. 
Similar to the case of three-spacetime dimensions
we find a narrow region in which the effective potential has 
a shape of first order phase transitions.
In the open inflation scenario a bubble nucleation (first order transition) 
occurs during the inflation \cite{Open}. It may be of interest to 
consider the possibility whether our model
works as a successful model for the open universe or not.
However this first order phase transition would not work successfully,
because the curvature (Hubble parameter) of de Sitter spacetime
changes rather fast during the inflation as we have shown in section 5.

We have briefly discussed a cosmological application in section 5.
We consider the phase transition in the chaotic inflation background,
neglecting the back-reaction from the fields which derives the phase 
transition. The strong coupling constant $\lambda$ leads to the early phase 
transition and the subsequent inflation dilute the strings.
The weak coupling constant does not lead to the phase transition 
during the inflation. Thus this model may work as a model of forming
the cosmic strings during the inflation if we choose the parameters 
in a suitable range \cite{Yokoyama,Lyth}.

\acknowledgments
The authors would like to thank Atsushi Higuchi, Roberto Camporesi, and 
Tomohiro Inagaki, for enlightening discussions and useful correspondences. 
We also thank Jun'ichi Yokoyama for useful comments and instructions. 
We are also grateful to Misao Sasaki, Michiyasu Nagasawa, 
and Koukichi Konno for useful conversations on this topics. 
One of the authors (T.~M.) is indebted to Monbusho Fund for
a financial support (No.08640377,11640280). The research by K.~Y. 
was partially supported by Inamori Foundation.

\newpage

\newpage

\begin{figure}[t]
  \begin{center}
  \leavevmode\psfig{file=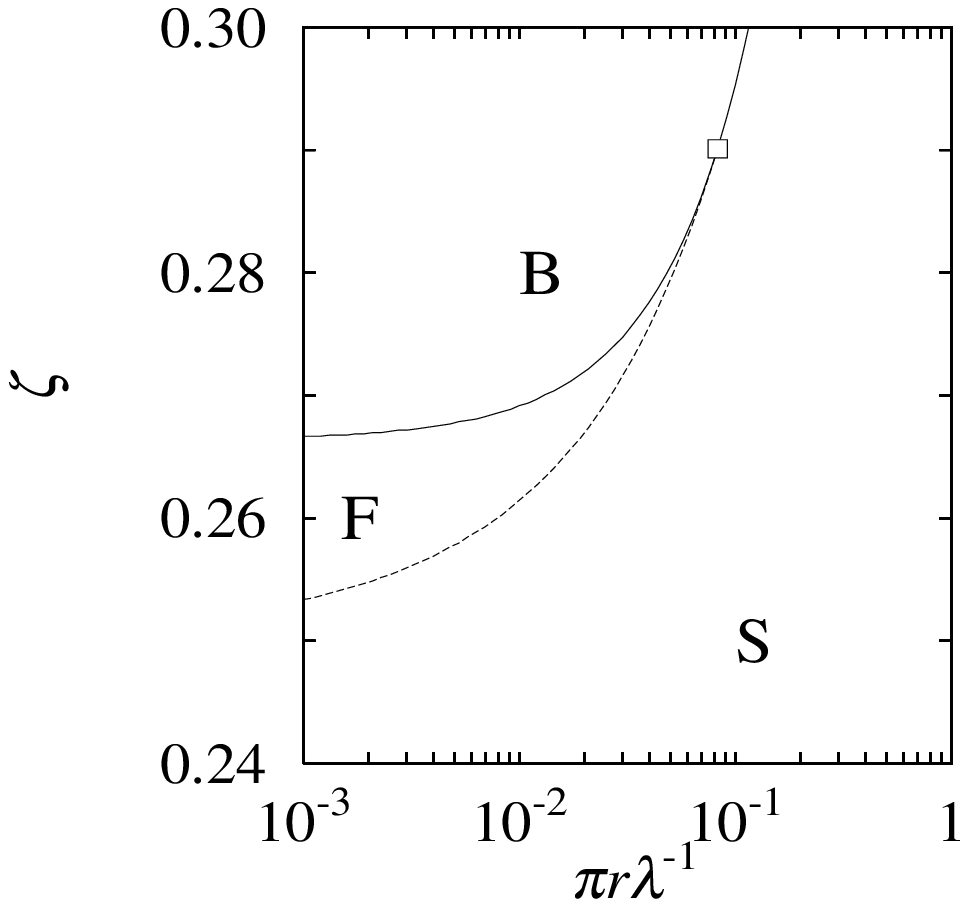,width=15cm}
  \end{center}
\caption{Phase structure of the model for $D=3$ on the 
$(\pi r/\lambda,\zeta)$ plane. 
In the regions labeled by 'S', 'B', and 'F', the
shape of the effective potential is of type of
unbroken symmetry, of second order phase transtion, and
of first order phase transition, respectively.
The square denotes the branching point.}
\label{Fig.pdthree}
\end{figure}

\begin{figure}[t]
  \begin{center}
  \leavevmode\psfig{file=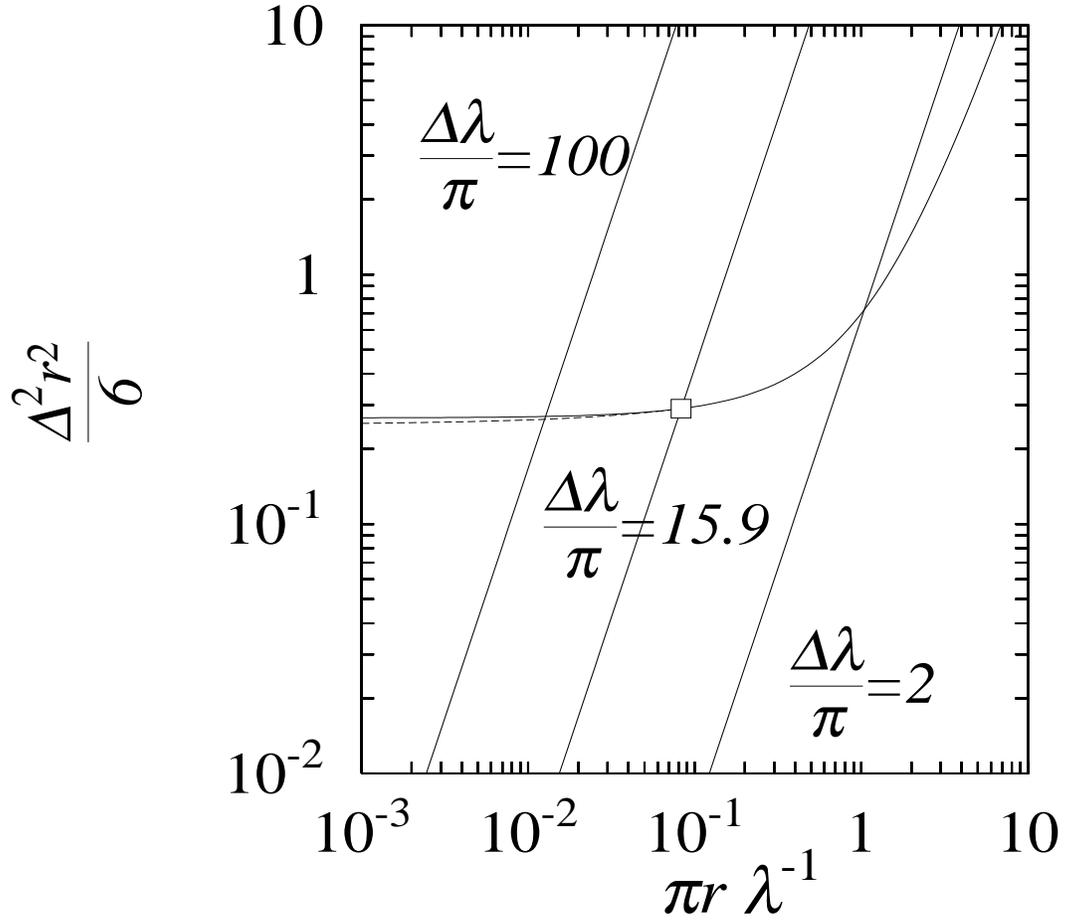,width=15cm}
  \end{center}
\caption{Typical trajectories for varying $r$ with $\Delta\lambda/\pi$
fixed on the $(\pi r/\lambda,\zeta)$ plane. We chose 
$\Delta\lambda/\pi=100$, ~$15.9$, and $2$ for straight lines, respectively.}
\label{Fig.arr}
\end{figure}

\begin{figure}[t]
  \begin{center}
  \leavevmode\psfig{file=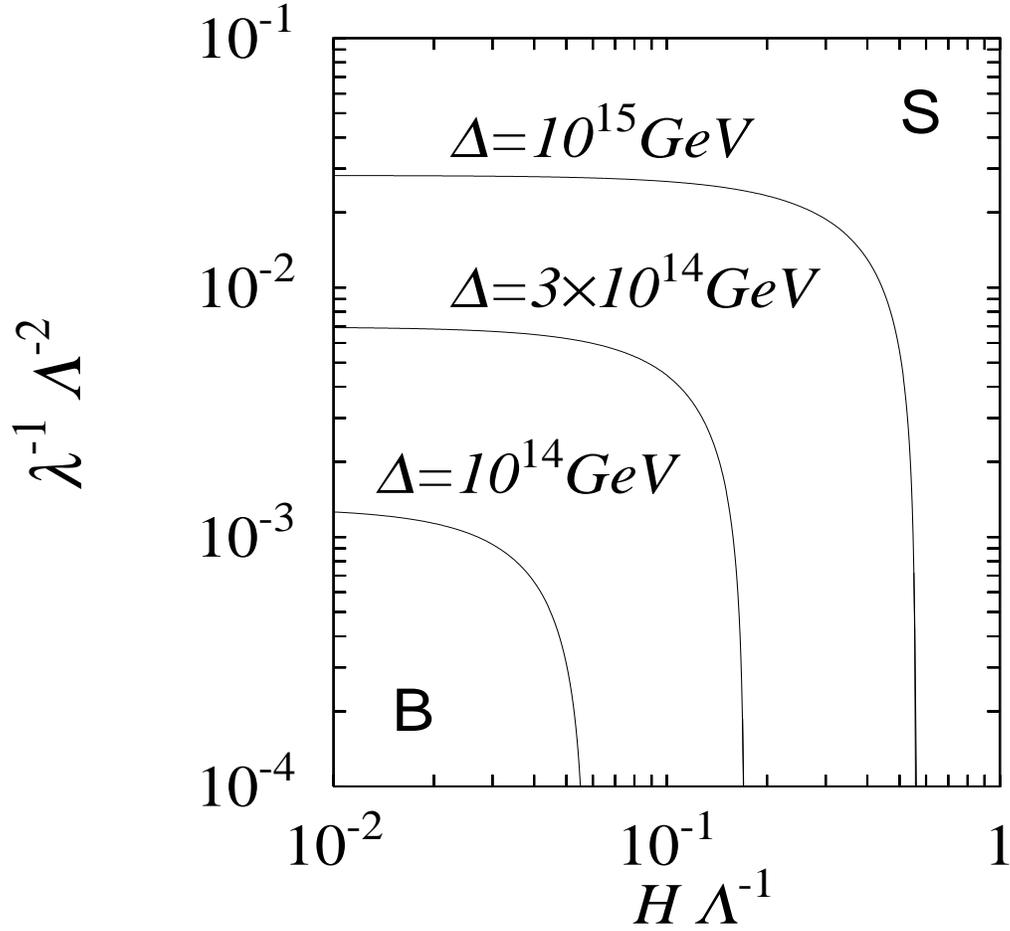,width=15cm}
  \end{center}
\caption{Phase boundary in the case $D=4$ on the 
$(H/\Lambda,1/\lambda\Lambda^2)$ plane. 
Here $\Lambda=10^{15}{\rm GeV}$ is used. 
We adopted the supersymmetry  breaking parameter 
$\Delta=10^{15},~3\times10^{14},~10^{14} ({\rm GeV})$,
respectively, for each curve in the figure. 
The right and upper region labeled by 'S' is
in a symmetric phase for each curve.
While the left and lower region labeled by 'B' is in a broken symmetry.}
\label{Fig.pd}
\end{figure}

\noindent
\begin{figure}[t]
  \begin{center}
  \leavevmode\psfig{file=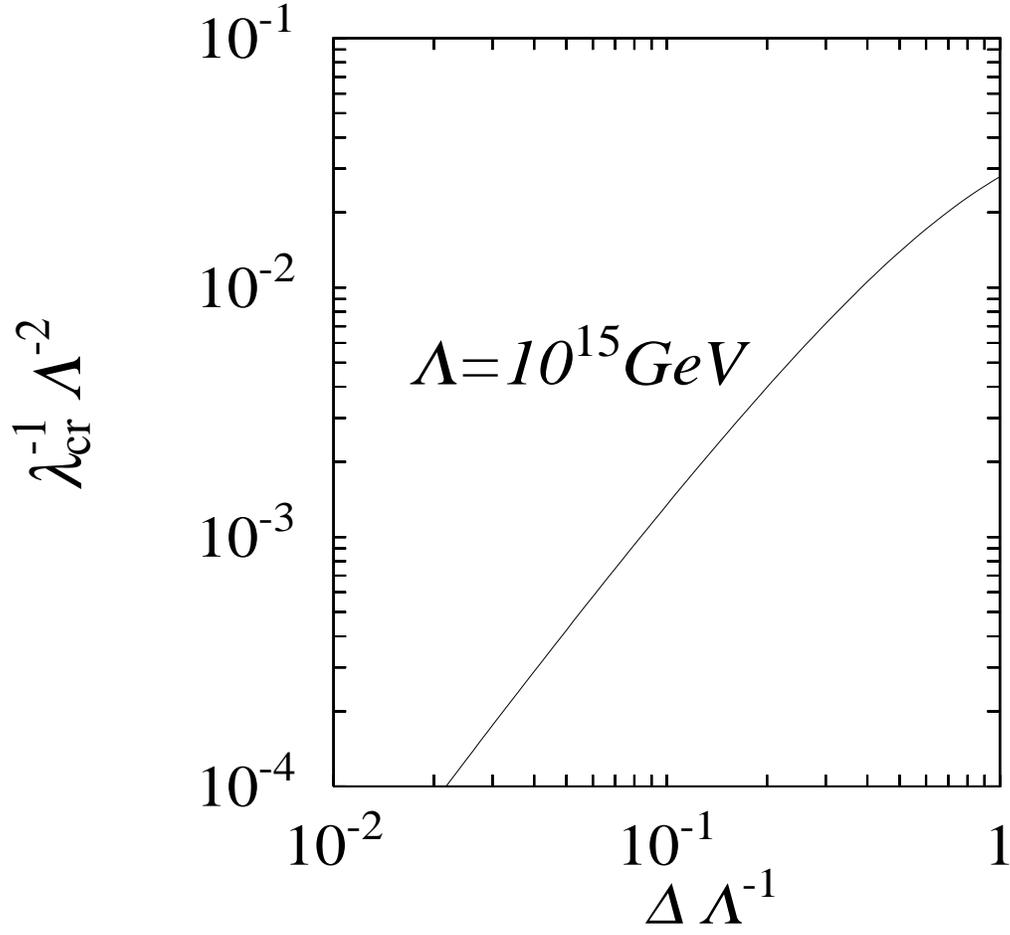,width=15cm}
  \end{center}
\caption{Critical coupling constant $\lambda_{\rm cr}^{-1}\Lambda^{-2}$ 
of equation (4.8) as a function of $\Delta/\Lambda$.}
\label{Fig.pdl}
\end{figure}

\begin{figure}[t]
  \begin{center}
  \leavevmode\psfig{file=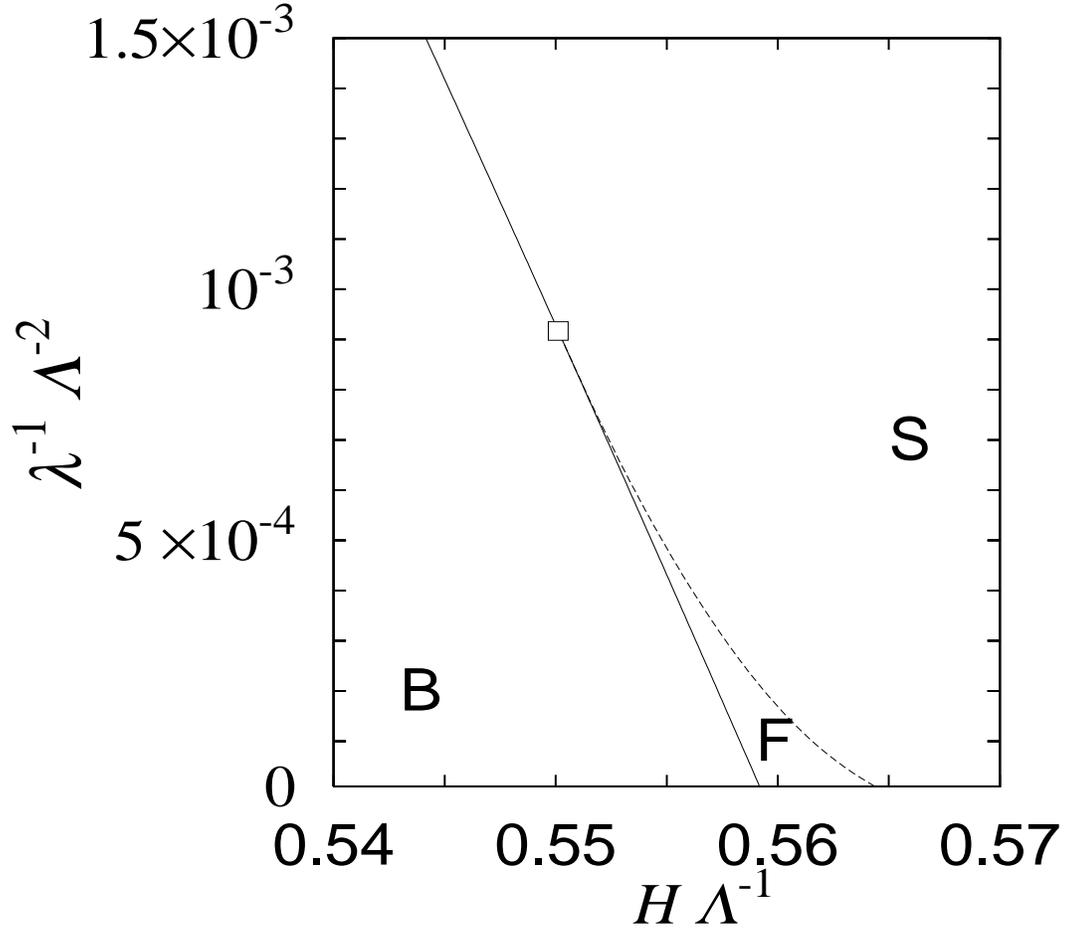,width=15cm}
\end{center}
\caption{Phase boundary for the model for $D=4$ on the 
$(H/\Lambda,1/\lambda\Lambda^2)$ plane. Here we set 
$\Lambda=10^{15}{\rm GeV}$ and $\Delta=10^{15}{\rm GeV}$.
In the narrow region between two curves labeled by $F$,
the effective potential has a shape of type of first order 
transition. The square is the branching point
obtained from numerical calculation.}
\label{Fig.pdf}
\end{figure}

\begin{figure}[t]
  \begin{center}
  \leavevmode\psfig{file=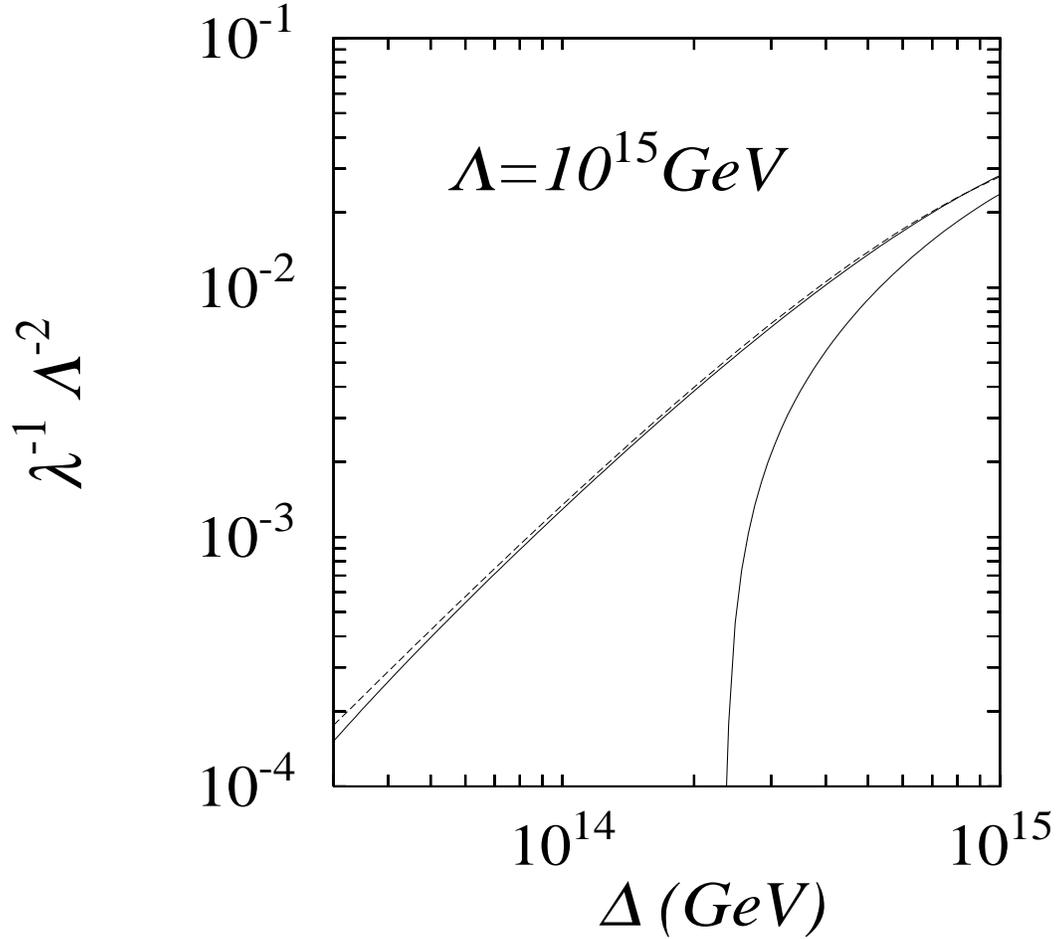,width=15cm}
  \end{center}
\caption{
Constraint on formation epoch of cosmic strings on the 
$(1/\lambda$, $\Delta)$ plane. Here we adopted
the momentum cutoff $\Lambda=10^{15}{\rm GeV}$.
The upper solid curve reflects the condition $\phi_2=(4\pi)^{-1/2}\Mpl$
and the lower solid curve corresponds to $\phi_2=3\Mpl$.
The dashed line shows the critical coupling constant $\lambda_{\rm cr}$.
}
\label{Fig.rnglmd}
\end{figure}

\end{document}